\documentclass[twocolumn]{jpsj3}
\usepackage{txfonts}

\usepackage{bm}
\begin{document}

\title{Ten-million-atom electronic structure calculations on the K computer \\
with a massively parallel order-$N$ theory %(DRAFT at. 2012.10.02 rev3)
%(DRAFT)
} 
\author{Takeo \textsc{Hoshi}$^{1,2}$
\thanks{Present e-mail address: hoshi@damp.tottori-u.ac.jp},
Yohei \textsc{Akiyama}$^1$,
Tatsunori \textsc{Tanaka}$^{1}$
\thanks{Present affiliation:MITSUBISHI Electric Control Software Corporation} and
Takahisa \textsc{Ohno}$^{3}$
}
\inst{
$^1$ Department of Applied Mathematics and Physics,
Tottori University, 4-101 Koyama-Minami, Tottori 680-8550, Japan \\
$^2$ Core Research for Evolutional Science and Technology,
Japan Science and Technology Corporation (CREST-JST), Sanbancho Bldg., 5, Sanbancho, Chiyoda-ku, Tokyo 102-0075, Japan \\
$^3$ National Institute for Materials Science, 1-2-1 Sengen, Tsukuba, Ibaraki 305-0047,  Japan
}

%%%%%%%%%%%%%%%%%%%%%%%%%%%%%%%%%%%%%%%%%
%
\abst{
A massively parallel order-$N$ electronic structure theory
was constructed by an interdisciplinary research 
between physics, applied mathematics and computer science.
(1) A high parallel efficiency
with ten-million-atom nanomaterials 
was realized on the K computer with upto 98,304 processor cores. 
The mathematical foundation is 
a novel linear algebraic algorithm for 
the generalized shifted linear equation.
The calculation was carried out by our code
\lq ELSES ' (www.elses.jp) with modelled (tight-binding-form) systems 
based on {\it ab initio} calculations. 
(2) A post-calculation analysis method, 
called $\pi$-orbital crystalline orbital Hamiltonian population ($\pi$-COHP) method,
is presented, since 
the method is ideal for huge electronic structure data distributed among massive nodes. 
The analysis method 
is demonstrated in an  sp$^2$-sp$^3$ nano-composite carbon solid,
with an original visualization software \lq VisBAR'. 
The present research indicates
general aspects of computational physics with current or next-generation supercomputers.
}

\kword{
order-N electronic structure theory,
numerical linear algebra, 
massive parallel computation, 
sp$^2$-sp$^3$ nano-composite carbon solid, 
crystalline orbital Hamiltonian population.
}

\date{\today}
\maketitle

\newcommand{\STACK}[2]{\genfrac{}{}{0pt}{1}{#1}{#2}}

A common issue in  current computational physics
is the theory for a large calculation with 
modern massively parallel supercomputers, like the K computer. 
A large calculation  
should accompany a large-data analysis theory,
as a post-calculation tool, 
so as to obtain a physical conclusion 
from huge numerical data distributed among massive computer nodes.
Interdisciplinary researches 
between physics, applied mathematics and computer science 
are crucial in this field and are sometimes called
\lq Application-Algorithm-Architecture co-design'.
%----------------------------------------------------------------------

The present paper is devoted to the theories, 
both for large-scale calculation and large-data analysis, 
as order-$N$ electronic structure theories
suitable for modern massively parallelized computers.

Order-$N$ electronic structure theories 
are those in which the computational time is proportional 
to the number of atoms in the system $N$
and are promising methods for large-scale calculation. 
The reference list can be found, for example, 
in Ref.~\cite{TENG-2011}.
Recently, 
several novel linear algebraic algorithms, with Krylov subspace, 
were developed for the order-$N$ theory,
{\it i. e.} 
generalized shifted conjugate-orthogonal conjugate-gradient method,
\cite{SOGABE-2009-CONF, TENG-2011, SOGABE-2012-GSQMR} 
generalized Lanczos method, \cite{TENG-2011} 
generalized Arnoldi method, \cite{TENG-2011} 
Arnoldi ($M,W,G$) method, \cite{YAMASHIT-2011-ArnoldiMWG}
multiple Arnoldi method, \cite{HOSHI-mArnoldi} 
generalized shifted quasi-minimal-residual method.
\cite{SOGABE-2012-GSQMR}
Their common foundation is 
the \lq generalized shifted linear equation', 
or the set of  linear equations 
\begin{eqnarray}
 ( z S -H ) \bm{x} = \bm{b}.
 \label{EQ-SHIFT-EQ}
\end{eqnarray}
Here $z$ is a (complex) energy value and 
the Hamiltonian and overlap matrices are denoted as $H$ and $S$
in the linear-combination-of-atomic-orbital (LCAO) representation, respectively.
They are sparse real-symmetric $M \times M$ matrices
and $S$ is positive definite.  
The vector $\bm{b}$ is an input and the vector $\bm{x}$ is the solution vector. 
Equation (\ref{EQ-SHIFT-EQ}) is solved, iteratively, 
instead of the generalized eigen-value equation 
($H  \bm{y}_k = \varepsilon_k S \bm{y}_k$).
The method is purely mathematical and 
may be useful also in other physics fields.
Researches in  the case of $S=I$ are found, for examples, 
in Quantum Chromodynamics \cite{FROMMER} and
many-body electron theory. \cite{YAMAMOTO-SCOCG} 

The multiple Arnoldi method is used in this paper,
since it is suitable for a molecular-dynamics (MD) simulation,
or the calculation of energy and force. \cite{HOSHI-mArnoldi}
See the paper~\cite{HOSHI-mArnoldi} for details. 
In short, 
the Green's function $G \equiv (zS-H)^{-1}$ is generated
in the LCAO representation 
and the computation is parallelized, in the most procedures of the code, 
by the column index $j$ of the Green's function ($G_{ij}$).
\cite{HOSHI-mArnoldi}
The physical quantities, such as energy and force, are calculated
through the one-body density matrix defined by
\begin{eqnarray}
 \rho_{ij} =  \frac{-1}{\pi} 
 \int f\left(\varepsilon - \mu \right) 
 {\rm Im} \, G_{ij}(\varepsilon + {\rm i} 0 ) \, d \varepsilon,
 \label{EQ-DM} 
 \end{eqnarray}
where 
$f$ is the occupation number in the Fermi-Dirac function 
with the chemical potential $\mu$. 
The method is applicable both to insulators and metals.
The method was implemented in a simulation package 
\lq ELSES' (www.elses.jp) 
with modelled or tight-binding(TB)-form Hamiltonians 
based on {\it ab initio} calculations. 
Several calculations were carried out 
with the charge-self-consistent formulation.
\cite{ELSTNER-1998-CSC}

\begin{figure}
    \begin{center}
      \includegraphics[width=8.5cm]{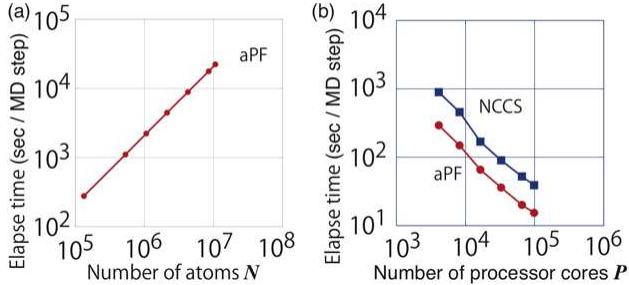}
\caption{
(a) Benchmark for the order-$N$ scaling. \cite{HOSHI-mArnoldi}
(b) Benchmark for the parallel efficiency (strong scaling) with $10^7$ atoms 
on the K computer among $P$ = 4,096 - 98,304 processor cores.
See the text for details.
}
    \label{FIG-BENCH}
  \end{center}
\end{figure}

In a previous paper, \cite{HOSHI-mArnoldi}
the order-$N$ scaling with upto ten-million atoms 
was confirmed, 
as shown in Fig. \ref{FIG-BENCH} (a)
and a parallel calculation was carried out
with 10$^3$ cores of Intel Xeon processors. 
The present paper will report
a calculation with 10$^5$ processor cores on the K computer 
and one will find several computational issues
clarified in 10$^5$-core calculation.

%============ Fig. 1 ===============================
%============ Fig. 1 ===============================

In the present paper, 
the parallel efficiency is measured on the K computer
for given materials with ten-million atoms. 
The resultant benchmark is called 
\lq strong scaling' in the high-performance computation society.
The numbers of the used processor cores are 
from $P=P_{\rm min} \equiv 4,096$ to $P=P_{\rm max} \equiv 98,304$. 
The MPI/OpenMP hierarchical parallelism was used and
the parallel unit for the MPI or OpenMP parallelism
is called \lq node' or \lq thread', respectively, throughout the present paper.
The number of threads is fixed to be $P^{\rm (thread)} = 8$,
the maximum value on the K computer and
the number of nodes is given by 
$P^{\rm (node)} = P / P^{\rm (thread)} = P /8$.
The calculations were carried out in a couple of MD steps, 
with a TB-form Hamiltonian, \cite{ASED-CALZAFFERI} 
for amorphous-like conjugated polymer, poly-(9,9 dioctil-fluorene) (aPF) 
with $N$=10,629,120 atoms \cite{HOSHI-mArnoldi} and 
sp$^2$-sp$^3$ nano-composite carbon solid (NCCS) 
with $N$=10,321,920 atoms
(See Ref.~\cite{HOSHI-2010-NPD} for a related study). 

As a result, 
a high parallel efficiency is found in Fig.~\ref{FIG-BENCH} (b). 
The measure for the efficiency is obtained as 
$\alpha \equiv T(P_{\rm min})/T(P_{\rm max}) \times (P_{\rm max}/P_{\rm min})=0.79$
or $\alpha=0.95$ for the aPF or NCCS case, respectively,  
where $T(P)$ is the elapse time with $P$ cores per MD step.
In general, 
the elapse time increases with the increase 
of the number of orbitals per atom ($m_{\rm orb}$), 
as well as the number of atoms ($N$).
The elapse time at the maximum cores is
$T(P_{\rm max})$=39.1 sec for the NCC system and
$T(P_{\rm max})$=15.3 sec for the aPF system,
since the NCCS system has four (s- and p-type) orbitals per atom ($m_{\rm orb}=4$)
and the aPF system contains hydrogen atoms with single (s-type) orbital 
and has, on average, only 2.3 orbitals per atom ($m_{\rm orb}=2.3$). 
The NCCS system was calculated also in 
a TB-form Hamiltonian with additional d orbitals 
($m_{\rm orb}=9$). \cite{CERDA}
The resultant elapse time  is 
$T(P_{\rm max})$=311.0 sec and 
is larger than that without the d orbital ($T(P_{\rm max})$=39.1 sec),
as should be.

Here several computational issues are discussed
for (I) data-communication saving
(II) memory-size saving and (III) parallel file reading/writing,
since they are crucial for a high performance
with practical computational resources. 
These issues affect the elapsed time and/or the required memory size
but does not affect the resultant values of physical quantities.
(I) The inter-node data communication is suppressed
by the redundant calculation of several quantities,
such as the matrix elements of $H$ and $S$, 
among the nodes. \cite{HOSHI-mArnoldi}
(II) A workflow for the memory saving was built in the code 
and the workflow saves the required memory size drastically, 
at the sacrifice of a moderate increase of the time cost. 
In the memory-saving workflow,  
the data array for the Green's function ($G$), the largest data array,  
is not stored in the memory but calculated twice redundantly, as follows:
\begin{eqnarray}
& & (H,S) \Rightarrow G \Rightarrow \mu \Rightarrow ({\rm 2nd \, calc. \, of \,}G)
\Rightarrow \rho.
 \label{EQ-WORK-FLOW}
\end{eqnarray}
The first calculation of the Green's function is carried out 
before the determination of the chemical potential $\mu$ in the bisection method.
After that, the second calculation is carried out,
since the density matrix $\rho$ is generated, as in Eq.~(\ref{EQ-DM}),  
both from the Green's function  $G$ and the chemical potential $\mu$. 
In a result with $10^5$ atoms by a single-node workstation, 
the consumed memory size is reduced drastically
from 28 GB in the non-memory-saving workflow 
into 1.6 GB in the memory-saving workflow,
while the increase of the time is moderate (19 \%). 
The reduction of the required memory is important,
since the built-in memory size of the K computer is only 16 GB per node.
All the calculations in Fig.~\ref{FIG-BENCH} 
were carried out in the memory-saving workflow. 
(III) Out test calculation shows that 
the parallel file reading can give an important acceleration
and was realized with split input files.
A typical file size with $10^7$ atoms is one G byte (B) 
for our input atomic-structure data
described in the extensible markup language (XML) format. \cite{NOTE-DATA-SIZE}
When the file with $N$ atoms is split into $n_{\rm split}$ files,
each split file contains the data with approximately $N/n_{\rm split}$ atoms.
The file writing is also parallelized,
when each node saves the atomic structure data in the split XML format. 
As a result with 4,096 cores (512 nodes), 
the consumed time for the initial procedures $T_{\rm ini}$,
or the procedures before starting the electronic structure calculation,
is drastically reduced,
from $T_{\rm ini}= 1,423$ sec with the non-parallel file reading
into $T_{\rm ini}= 69$ sec with the parallel file reading. 
\cite{NOTE-FILE-READ}
The file writing procedure consumes a tiny time cost ( 0.2 - 0.4 sec).
It is noteworthy that
in most MD simulation studies,
the file writing procedure of the atomic structure 
is carried out 
with a certain interval of MD steps, typically 10 steps.
Therefore, 
the file writing time will be negligible 
in the whole simulation time.

%%%%%%%%%%%%%%%%%%%%%%%%%%%%%%%%%%%%%%%%%%%%%%%%%%%%%%%%%
%\section{Green-function based analysis}\label{sec_cohp}

%============ Fig. 2 ===============================
\begin{figure*}[ht]
    \begin{center}
      \includegraphics[width=19cm]{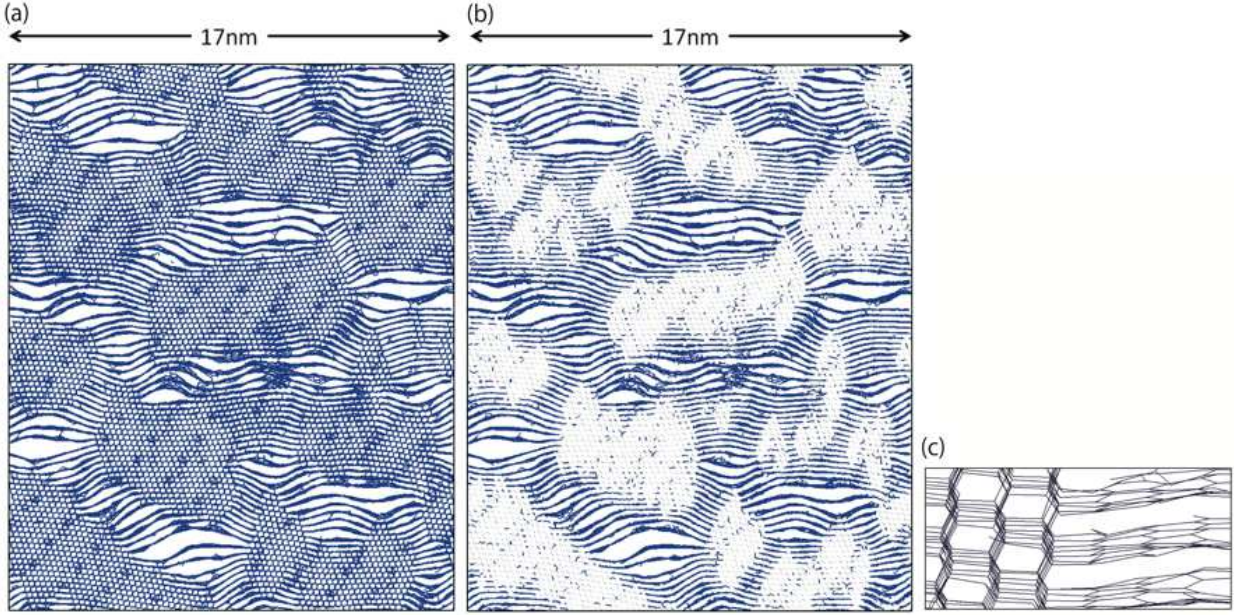}
\caption{
Bond visualization of sp$^2$-sp$^3$ nano-composite carbon solid (NCCS)
with 10$^5$ atoms by the COHP and $\pi$COHP analysis. 
The bonds are visualized both for the sp$^2$ and sp$^3$ domains,
or both for $\sigma$ and $\pi$ bonds in (a),
while the bonds are visualized only for the sp$^2$ domain
or for $\pi$ bonds in (b). 
A closeup of a sp$^2$-sp$^3$ domain boundary is shown in (c). 
}
    \label{FIG-NPD-COHP}
  \end{center}
\end{figure*}
%============ Fig. 2 ===============================

Now the discussion is turned into the second topic, 
the post-calculation analysis with huge electronic structure data. 

The present paper presents analysis methods based on the Green's function,
{\it i. e.} 
crystalline orbital Hamiltonian population (COHP) method, 
\cite{COHP-1993} and its theoretical extension called $\pi$COHP method. 
The original COHP reveals the local bonding nature for each atom pair energetically
and its definition is  
\begin{eqnarray}
 C_{IJ}(\varepsilon)  \equiv \frac{-1}{\pi}  \sum_{\alpha, \beta}  
H_{J \beta; I \alpha} {\rm Im} \, G_{I \alpha; J \beta}(\varepsilon + {\rm i} 0 ),
 \label{EQ-COHP} 
\end{eqnarray}
where the basis suffices, previously denoted by $i$ and $j$, 
are decomposed into the atom suffices, $I$ or $J$, 
and the orbital suffices, $\alpha$ or $\beta$ 
($i \equiv (I, \alpha), j \equiv (J, \beta)$).
The energy integration of COHP, called ICOHP, is also defined as
\begin{eqnarray}
 B_{IJ} \equiv \int f\left(\varepsilon - \mu \right)  C_{IJ}(\varepsilon) d \varepsilon.
 \label{EQ-ICOHP} 
\end{eqnarray}
The sum of ICOHP gives
the electronic structure energy or the sum of the occupied eigen levels; \cite{COHP-1993}
\begin{eqnarray}
 E_{\rm elec}  \equiv \sum_k f(\varepsilon_k - \mu ) \, \varepsilon_k = \sum_{I,J} B_{IJ}.
 \label{EQ-ICOHP-SUM} 
\end{eqnarray}
A large negative value of $B_{IJ}$ indicates
the energy gain for the bond formation between the ($I,J$) atom pair. 
In the original paper, \cite{COHP-1993} 
the analysis is carried out for {\it ab initio} calculations 
in the linear-muffin-tin-orbital formulation \cite{LMTO-1984}.
The method was applied not only to insulator
but also to metal, such as in Ref.~\cite{Y_ISHII-QC-COHP}. 
The method is suitable for the present order-$N$ method, 
since the present method is based on the Green's function.  
\cite{TAKAYAMA-KRSD-COCG}

Here the $\pi$COHP is proposed as
a theoretical extension of the original COHP.
If the Hamiltonian contains the s- and p-type orbitals, for example, 
the off-site Hamiltonian term is decomposed 
into the $\sigma$ and $\pi$ components 
($H_{J \beta; I \alpha} = H_{J \beta; I \alpha}^{(\sigma)} + 
H_{J \beta; I \alpha}^{(\pi)})$. 
The $\pi$COHP, $C_{IJ}^{(\pi)}$, will be defined in Eq.~(\ref{EQ-COHP}),
when $H_{J \beta; I \alpha}$ is replaced by $H_{J \beta; I \alpha}^{(\pi)}$.
The $\pi$ICOHP, $B_{IJ}^{(\pi)}$, will be defined in Eq.~(\ref{EQ-ICOHP}), 
when $C_{IJ}$ is replaced by $C_{IJ}^{(\pi)}$.
The $\sigma$COHP, $C_{IJ}^{(\sigma)}$, and 
the $\sigma$ICOHP, $B_{IJ}^{(\sigma)}$, are defined in the same manners. 
From the definitions, 
the (I)COHP is decomposed into the sum of the $\sigma$(I)COHP and $\pi$(I)COHP
\begin{eqnarray}
  C_{IJ}(\varepsilon) &=& C_{IJ}^{(\sigma)}(\varepsilon) + C_{IJ}^{(\pi)}(\varepsilon)
   \label{EQ-COHP-DECOMPOSE} \\
  B_{IJ} &=& B_{IJ}^{(\sigma)} + B_{IJ}^{(\pi)}.
    \label{EQ-ICOHP-DECOMPOSE} 
\end{eqnarray}
Hereafter the original (I)COHP is called as \lq full' (I)COHP. 
In the code, 
the full, $\sigma$ and $\pi$(I)COHP can be calculated automatically, 
without any additional data communication, 
during the massively parallelized order-$N$ calculation.
One can distinguish the $\sigma$ and $\pi$ bonds, 
energetically, by the $\sigma$(I)COHP and $\pi$(I)COHP.
A large negative value of $B_{IJ}^{(\sigma)}$ or $B_{IJ}^{(\pi)}$ indicates
the $\sigma$ or $\pi$ bond formation between the atom pair, respectively. 

In Fig.~\ref{FIG-NPD-COHP}, 
the ($\pi$)COHP analysis is demonstrated 
in an NCCS system with 10$^5$ atoms,
so as to distinguish the sp$^2$ and sp$^3$ domains,
because 
one can distinguish the sp$^2$ domains
from the sp$^3$ domains
by the presence of $\pi$ bonds. 
The system is a resultant structure 
of the our previous finite-temperature MD simulation 
with a periodic boundary condition. 
\cite{HOSHI-2010-NPD}
The simulation is a preliminary research for the formation process of 
the nano-polycrystalline diamond (NPD), a novel ultra-hard materials. 
\cite{IRIFUNE-2003-NPD, GUILLOU-2007-NPD-GROWTH}
The NPD is produced directly from graphitic materials
and consists of 10-nm-scale diamond-structure domains
with a characteristic lamella-like structure. 
Its growth process is of great interest and 
possible precursor structures should be 
a ten-nm-scale composite 
between the sp$^2$ (graphite) and sp$^3$ (diamond) domains. 
The present research is motivated from the above problem,
though the present structures, still, 
have a gap in the length scale,
when it is compared with experiments,
as discussed later. 

Figure \ref{FIG-NPD-COHP} (a) or (b) shows 
the bond visualization 
with the full ICOHP or the $\pi$ICOHP, respectively.
In Fig.~\ref{FIG-NPD-COHP}(a), 
a bond is drawn for an $(I,J)$ atom pair, 
when its ICOHP value satisfy the condition of 
$B_{IJ} < B_{\rm th}$
with a given threshold value of $B_{\rm th} (<0)$.
We found a typical value of $B_{\rm th} = - 9$ eV.
The visualization with the full ICOHP indicates 
the visualization both for sp$^2$ and sp$^3$ bonds. 
In Fig.~\ref{FIG-NPD-COHP}(b), on the other hand, 
a $\pi$ bond is drawn,   
when its $\pi$ ICOHP value satisfy the condition of 
$B_{IJ}^{(\pi)} < B_{\rm th}^{(\pi)}$
with a given threshold value of $B_{\rm th}^{(\pi)} (<0)$.
We found a typical value of $B_{\rm th}^{(\pi)} = - 1.5$ eV.
The visualization with the $\pi$ICOHP indicates 
the visualization only for sp$^2$ bonds.

The bond visualization analysis in Figs.~\ref{FIG-NPD-COHP}(a) and (b) 
concludes that the layered domains are sp$^2$ domains
and the non-layered domains are sp$^3$ domains. 
Figure \ref{FIG-NPD-COHP}(c) shows 
the visualization of a boundary region 
between sp$^2$ and sp$^3$ domains.
Here one can confirms that a layered domain 
form an sp$^2$ or graphite-like structure 
and a non-layered domain form an sp$^3$ or diamond-like structure,
as expected from the ($\pi$)ICOHP analysis. 

 Several points are discussed for the ($\pi$)COHP analysis. 
(I) The value of $| B_{\rm th}^{(\pi)} |$
is much smaller than that of $| B_{\rm th}|$, 
because the $\pi$ bonding is 
much weaker than the $\sigma$ bonding. 
(II) The threshold values for the bond visualizations, 
$B_{{\rm th}}$ and $B_{{\rm th}}^{(\pi)}$,  
may not be universal among materials
but is independent on the system size.
One should choose a typical value once for a material
and can use the value among different system sizes. 
(III) One should be careful, sometimes, in the interpretation of the analysis result,
because the $\pi$COHP analysis dose  not detect an sp$^2$ bond 
but detects a contribution of the $\pi$-bonding component, as explained above. 
For example, 
the initial structure for the simulation of Fig.~\ref{FIG-NPD-COHP} 
contains defects among the sp$^3$ domains, \cite{HOSHI-2010-NPD}
as \lq seeds' of the sp$^2$-sp$^3$ domain boundary. 
The resultant structure in Fig.~\ref{FIG-NPD-COHP} still has
several initial defects in the sp$^3$ domains
and $\pi$ bonds are often drawn in Fig.~\ref{FIG-NPD-COHP}(b) 
at such local defective regions. 
Such a $\pi$ bond does not mean an $sp^2$ bond.
(IV) The  structure of Fig.~\ref{FIG-NPD-COHP} 
has a gap in the length scale,
when it is compared with experiments.
The structure is a \lq 2D-like' one,
because the periodic cell length 
in the perpendicular direction to the paper (2nm) 
is much smaller than the other two cell lengths, 17 nm or more. \cite{HOSHI-2010-NPD}
The artificial \lq 2D-like' situation affect severely 
the resultant atomic structure and 
makes a difficulty for a direct comparison between the simulation and experiment.
A more realistic situation with the ten-nanometer simulation cell sizes 
in the three directions
requires million-atom MD simulation, 
ten times larger than that in the present result.
\cite{HOSHI-2010-NPD}
Such a larger MD simulation
may be a possible target in near future with the parallel computations.  
(V) The bond visualization of 
Figs.~\ref{FIG-NPD-COHP}(a)-(c)
was realized by our original visualization tool 
'VisBAR'(=Visualization tool with Ball, Arrow and Rods). 
The tool is based on Python (www.python.org) and 
was developed for our needs 
in the large-scale calculations,
like the ($\pi$)COHP analysis.

In summary,
(i) a high parallel efficiency was found in 
ten-million-atom order-$N$ electronic structure calculations
on the K computer with approximately $10^5$ processor cores. 
Important computational issues are addressed
for communication, memory size and file reading/writing.
(ii) The ($\pi$) COHP analysis method is presented 
as a practical post-calculation analysis method
ideal for the huge distributed data of the Green's function.
The analysis is demonstrated in a sp$^2$-sp$^3$ nano-composite carbon solid,
so as to distinguish sp$^2$ and sp$^3$ domains.
The example shows a typical need of 
large-scale electronic structure calculation
that requires
both large-scale calculation 
and large-data analysis with huge distributed data.

The present research indicates
general aspects of computational physics,
beyond electronic structure calculation,
with current or next-generation supercomputers.
Numerical algorithm and computer scientific methods
will be inseparable from physics. 
A physical discussion should be
constructed from physical quantities, 
like the Green's function or the ($\pi$-)COHP,  
that can be computed, analyzed and visualized
with massively parallel computer architectures.
All the methods discussed here, 
ones for calculation, 
parallel file reading/writing with split XML file,
memory saving workflow,  
post-calculation data analysis, and visualization, 
are designed to be suitable for the massively parallel computer architecture,
and some of them may be useful in other computational physics fields.

%%%%%%%%%%%%%%%%%%%%%%%%%%%%%%%%%%%%%%%%%%%%%%%%%%%%%%%%%
\begin{acknowledgment}

{\bf Acknowledgments} 
The present research is partially supported by the Field 4 
( Industrial Innovation )
of the HPCI Strategic Program of Japan. 
A part of the results is obtained 
by the K computer at the RIKEN Advanced Institute for Computational Science
(The early access and the proposal numbers of hp120170, hp120280).
This research is supported partially by Grant-in-Aid 
for Scientific Research
(No. 23104509, 23540370)
from the Ministry of Education, Culture, Sports, Science and Technology (MEXT) of Japan. 
This research is supported also partially by Initiative on Promotion of Supercomputing for Young Researchers, 
Supercomputing Division, Information Technology Center, The University of Tokyo.
The supercomputers were used also 
at the Institute for Solid State Physics, University of Tokyo
and at the Research Center for 
Computational Science, Okazaki. 

\end{acknowledgment}

%%%%%%%%%%%%%%%%%%%%%%%%%%%%%%%%%%%%%

%%%%%%%%%%%%%%%%%%%%%%%%%%%%%%%%%%%%%%%%%%%%%%%%%%%%%%%%%

\begin{thebibliography}{99}
\bibitem{NOTE-DATA-SIZE}
Since the atomic structure data
requires three ($x,y,z$) components in the double precision (8B) value for each atom,
the required data size with $10^7$ (=10M) atoms is estimated to be
$3 \times 8 {\rm B} \times 10$M = 240 MB.
\bibitem{NOTE-FILE-READ}
The consumed time only for the file reading can not be measured,
since other procedures,  
such as array allocation and calculations,
are carried out during the file reading,
\end{thebibliography}
\end{document}